\documentstyle[preprint,tighten,aps, prc]{revtex}

\newcommand{\resetcounter}{\setcounter{equation}{0}}  

\newcommand{\GG} {{\Delta}}               

\newcommand{\A}  {\alpha}

\newcommand{\p}{\partial}

\begin{document}


\draft
\preprint{\hfill\vbox{\hbox{UUPHY/97/10}
\hbox{physics/9710040}}}
\title{Nonlinear Dirac and diffusion equations in 1 + 1 
dimensions from stochastic considerations}
\author{Karmadeva Maharana}
\address{Department of Physics, Utkal University, Bhubaneswar- 751004,
India}
\maketitle\

\begin{abstract}
We generalize the method of obtaining the fundamental linear
partial differential equations such as the diffusion and 
Schr\"odinger
equation, Dirac and telegrapher's equation from a simple 
stochastic consideration to arrive at certain nonlinear 
form of these equations. The group classification through one
parameter group of transformation for two of these equations
is also carried out.    
\end{abstract}



%
%


\newpage



\section{Introduction}
\resetcounter

It is remarkable that some of the fundamental linear equations
of physics such as the diffusion and Schr\"odinger equations, Dirac
and telegrapher's equations, and the 
Maxwell equations can be obtained by setting up a master equation
from simple stochastic considerations and a modification there of
\cite{Gaveau,Ord,OrdG}. 

Hosts of nonlinear partial differential equations have been vigorously
studied in recent times for possessing interesting solitonic, 
self focussing and allied properties in fields wide ranging from
lasers to string theories. It is interesting to note, as it is
shown in classical mechanics with randomness playing a key role,
that any arbitrary initial distribution of velocities must eventually 
become Maxwellian. Thus the Maxwellian distribution must be invariant
under the underlying stochastic process \cite{Chandrasekhar}. 
The system tending 
to and remaining in Maxwellian distribution is analogous to the
self focussing in nonlinear optics and the properties of solitonic
systems arising in nonlinear phenomena.

The transition to nonlinear equation from a linear equation through 
certain transformations is well known. An example is the Cole-Hopf
transformation that carries over the linear diffusion equation to the 
nonlinear Burgers equation. The reverse process of getting linear
diffusion equation from a nonlinear diffusion equation in the form
$\frac{\p \phi}{\p t} = v^2 \frac{\p^2 \phi}{\p v^2}$ has also been
studied \cite{Munier,Hill} through a nonlinear transformation.

In this paper, we propose to obtain a class of nonlinear equations in a 
different way by generalizing the method of \cite{Gaveau}.  The method is 
simple. The basic inputs can be incorporated from considerations
and arguments based on physical reasoning to obtain nonlinear equations 
rather than arbitrary mathematical transformations. The form of the
equations obtained are quite restrictive. However, we do not address 
the deep mathematical significance of Cole-Hopf transformation and the 
like in this method.

In section II, we briefly review the method used in setting up some
basic linear equations of physics. Then we generalize the procedure
of obtaining classes of correspondng nonlinear partial differential
equations in section III. Next, section IV is devoted to the 
construction of the groups under which two of the equations obtained
in section II, namely diffusion equation with nonlinearity and nonlinear
telegrapher's equation, remain invariant. The similarity 
transformation and the 
Lie algebra are constructed to show the transformations under which 
solutions go over to new solutions. 

\section{Linear equations}
The fundamental role of linear diffusion equation in Physics and
its applications in other branches hardly needs any elaboration. 
An analytic continuation turns this equation into Schr\"odinger
equation. Another basic equation of physics, the Dirac equation 
originated in the attempt to make Schr\"odinger equation compatible
with special theory of relativity. It is a simple matter to arrive 
at telegrapher's equation by iterating Dirac equation in (1 + 1) 
dimensions. It is a curiuos fact that these equations have been 
obtained by Gaveau etal \cite{Gaveau} and Ord 
\cite{Ord} from stochastic consideration
by setting up a master equation. Ord \cite{OrdG} has also arrived at 
the Maxwell equation by a modification of the master equation. 

Following \cite{Gaveau,Ord,OrdG} we briefly review how these 
equations are 
achieved and then proceed to nonlinear generalization.
The basic consideration is the correlation over a random ensemble
of particles. However, for simpler visualization we may follow the 
Boltzmann approach by analysing the movement of a single particle. 
Let a particle have random motion in one space dimension moving with
a fixed speed $v$. We assume that it has complete reversal of direction
of motion in a random manner from time to time, say as to flip of 
a coin. So this is according to Poisson distribution, that is to say 
that there is a fixed rate $a$ for this reversal and the probability 
for reversal in a time interval $d t$ is $a dt$. Let $P_{+}(x, t)$
(respectively, $P_{-}(x, t)$) be the probability density for the 
particle being at $x$ at time $t$ and moving to the right (respectively
left). The master equation for an infinitesimal time step is 

\begin{eqnarray}
P_{\pm}(x, t + \Delta t) = P_{\pm} (x \mp \Delta x, t)(1 - 
a \Delta t) + P_{\mp}(x \pm \Delta x, t) a \Delta t 
\end{eqnarray}

This equation gives rise to the linear equations such as Dirac,
telegrapher's, diffusion or Schr\"odinger equations in the lowest
approximation under various circumstances.

To the lowest order in $\GG x$ and $\GG t$, equation (1) gives,
\begin{eqnarray}
\frac{\p P_{\pm}}{\p t} = - a (P_{\pm} - P_{\mp}) \mp v \frac{\p 
P_{\pm}}{\p x},
\hspace{1cm} v = \frac{\GG x}{\GG t}.
\end{eqnarray}
and the telegrapher's equation follows by iteration,
\begin{eqnarray}
\frac{\p^2 P_{\pm}}{\p t^2} - v^2 \frac{\p^2 P_{\pm}}{\p x^2} = 
- 2 a \frac {\p P_{\pm}}{\p t}.
\end{eqnarray}

The one dimensional Dirac equation is obtained from (1) by analytic 
continuation. First we identify $P_{\pm}$ with $u_{\pm}$, 
$v\leftrightarrow
c$, $\frac{i m c^2}{\hbar}\leftrightarrow a$, and then perform 
a phase transformation 
$u(x, t) = e^{\frac{i m c^2 t}{\hbar}} \Psi(x, t)$. This results in
\begin{eqnarray}
i \hbar \frac{\p\Psi}{\p t} = m c^2 \sigma_x \Psi - 
i c \hbar \sigma_z
\frac{\p \Psi}{\p x}.
\end{eqnarray}
  
In the Feynman path integral formulation through checkers moves
on space-time, 1 has to be replaced by a factor $1 + (\frac{i m c^2}
{\hbar}) \GG t$ for each step on which a reversal does not 
take place, whereas for reversals there is a factor $- i\GG t 
\frac{m c^2}{\hbar}$.

The Dirac equation in (1 + 1) dimensions, having 
two components, has the similar time and space dependence in 
this stochastic approach. But for a single component object we
find that the linear diffusion equation results, which shows the 
asymmetry in derivatives arising out of the random walk 
problem. 

A generalization to three space dimensions has been carried out
in \cite{Gaveau}. 

McKeon and Ord \cite{McKeon} have shown that if movements backward and 
forward in time is superposed as well on the previous motion, 
then Dirac equation in one dimension results without recourse to
direct analytic continuation. 

To arrive at the linear diffusion equation in a simple way, we 
put $P_{\pm} = P_{\mp} = P$, and $a = \frac{1}{2 \GG t}$. The master 
equation (1) reduces to 
\begin{eqnarray}
P(x, t + \GG t) = \frac{1}{2} P(x - \GG x, t) + \frac{1}{2}
P(x + \GG x, t).
\end{eqnarray}
Expanding this in a Taylor series about the point $(x, t)$ gives,
\begin{eqnarray}
P(x, t) + \frac{\p P(x, t)}{\p t} \GG t + \ldots = P(x, t) + 
\frac{\p^2 P(x, t)}{\p x^2} \frac{(\GG x)^2}{2} + \ldots
\end{eqnarray}
and equating the lowest order terms we get,
\begin{eqnarray}
\frac{\p P}{\p t} = \frac{\p^2 P}{\p x^2} \frac{(\GG x)^2}
{2 \GG t} = D \frac{\p^2 P}{\p x^2}
\end{eqnarray}
where, $D = \frac{(\GG x)^2}{2 \GG t}$. 

It may be noted that the above equation in the context of 
Brownian motion can be obtained from a consideration of one 
dimensional random walk with a Bernoulli distribution of 
probability and the statistical considerations sets \cite
{Chandrasekhar} 
\begin{eqnarray}
\lim_{\GG t\rightarrow 0} \frac{(\GG x)^2}{2 \GG t} = D
\end{eqnarray}
where $D$ is a constant. 

A formal analytic continuation 
({\it e. g.} $t\rightarrow i t$, or $D\rightarrow i \hbar$) 
leads to Schr\"odinger equation for free particles. A potential $V(x,t)$
can be included by adding a term $V(x,t) P(x,t)\GG t$ to the right hand
side of equation (5). 

Ord \cite{OrdG} has obtained the Maxwell equations in 1+1 dimensions by
a modification of the master equation. We follow his procedure to show
 how it is done.
First equation (1) is modified to,
\begin{eqnarray}
P_{\pm} (x, t + \GG t) = P_{\pm}(x\mp \GG x, t) + a (x, t) \GG t
\end{eqnarray}
where, $a(x, t)$ is interpreted as a source and linear combinations
of $P_+$ and $P_-$ will correspond to the potentials $A(x, t)$
and $\Phi(x, t)$. To the lowest order in $\GG x$ and $\GG t$ 
equation (9) gives,
\begin{eqnarray}
\frac{\p P_{\pm}(x, t)}{\p t} \GG t = \mp \frac{\p P_{\pm}}{\p x} 
\GG x + a(x, t) \GG t.
\end{eqnarray}
Writing,
\begin{eqnarray}
A (x, t) &=& \frac{1}{2} [P_+(x, t) + P_-(x, t)]
\nonumber\\ 
\Phi (x, t) &=& \frac{1}{2} [P_+(x, t) - P_-(x, t)]
\end{eqnarray}
equation (10) implies,
\begin{eqnarray}
\frac{\p A(x, t)}{\p t} &=& - c \frac{\p \Phi(x, t)}{\p x}
+ a(x, t)\\ 
\frac{\p \Phi(x, t)}{\p t} &=& -c \frac{\p A(x, t)}{\p x}
\end{eqnarray}
where we have put $\frac{\GG x}{\GG t} = c$.

Equations (10) and (11) may be decoupled by differentiating the
first with respect to $t$ and the second with respect to $x$ to 
give ,
\begin{eqnarray}
\frac{\p^2 A(x, t)}{\p t^2} = c^2 \frac{\p^2 A(x, t)}{\p x^2}
+ \frac{\p a(x, t)}{\p t}
\end{eqnarray}
and similarly we get,
\begin{eqnarray}
\frac{\p^2\Phi(x, t)}{\p t^2} = c^2 \frac{\p^2 \Phi(x, t)}
{\p x^2} - c \frac{\p a(x, t)}{\p t}
\end{eqnarray}
Equations (13), (14), and (15) are equivalent to Maxwell equations
in (1 + 1) dimensions, equation (13) being the Lorentz condition
\begin{eqnarray}
\frac{\p A(x, t)}{\p x} + \frac{1}{c} \frac{\p \Phi(x, t)}
{\p t} = 0.
\end{eqnarray}
In order to obtain the wave equation for the ``vector potential"
$A$, we write,
\begin{eqnarray}
\frac{1}{c} \frac{\p a(x, t)}{\p t} = 4 \pi J(x, t)
\end{eqnarray}
and equation (14) becomes,
\begin{eqnarray}
\frac{\p^2 A(x, t)}{\p x^2} - \frac{1}{c^2} \frac{\p^2 A(x, t)}
{\p t^2} = - \frac{4\pi}{c} J(x, t)
\end{eqnarray}
and similarly writing
\begin{eqnarray}
\frac{1}{c}\frac{\p a(x, t)}{\p x} = - 4\pi\rho (x, t)
\end{eqnarray}
equation (15) becomes the wave equation for scalar potential
$\Phi(x, t)$,
\begin{eqnarray}
\frac{\p^2 \Phi(x, t)}{\p x^2} - \frac{1}{c^2}\frac{\p^2\Phi(x, t)}
{\p t^2} = - 4 \pi \rho(x, t).
\end{eqnarray}
The two definitions (17) and (19) imply that
\begin{eqnarray}
\frac{\p J(x , t)}{\p x} + \frac{\p\rho(x, t)}{\p t} = 0
\end{eqnarray}
which is the equation of continuity. These considerations may be 
generalized to three space dimensions. 

The objective of the above long review is to stress the interesting 
fact that many of the fundamental linear equations of physics
are obtainable from an elementary consideration of stochastic 
process. Of course, by no stretch of imagination, we may expect
the whole of physics to follow from such a consideration. 

\section{Nonlinear equations}

Nonlinear partial differential equations appear in all branches
of physics 
and some of these have interesting properties such as soliton
like solutions, infinite number of conserved objects and so on. 
The nonlinear diffusion equation in the form (in our notation)
\begin{eqnarray}
\frac{\p P(x, t)}{\p t} = \frac{\p}{\p x} \left [f(p)\frac{\p P(x, t)}
{\p x}\right ]
\end{eqnarray}
is well known in literature 
\cite{Ovsiannikov,Bluman,Munier,Hill,Olver} and the properties of its 
solutions have been extensively studied. 

Now we proceed with an aim at getting the above nonlinear equation
and others out of the master equation (1), by suitable modifications.
If we consider this to be a phenomenological equation, without 
any recourse to Poisson's distribution, then the obvious way to 
introduce nonlinearity is to introduce functions of $x$ and $t$ 
as multiplicative coefficients on right hand side of equation (1). 
As an example, suppose we apply randomness to $\GG x$ itself. 
We know that if $\alpha$ is chosen at random in the interval 
(0, 1) then the probability of $\alpha_n$ to be in the 
interval $(x, x + d x)$ is given by $\frac{d x}{(1 + x)\log 2}$
for large $n$, and hence a possibility is to use $\frac{\GG x}
{(1 + x)\log 2}$ instead of $\GG x$ in equation (5), which is 
a special case of equation (1). Perhaps it would be simplest to 
replace $\GG x$ by $P(x, t) \GG x$ in (5), and the resulting equation
is 
\begin{eqnarray}
\frac{\p P(x, t)}{\p t} = D P^2(x, t) \frac{\p^2 P(x, t)}{\p x^2}
\end{eqnarray}
where
\begin{eqnarray}
D = \frac{(\GG x)^2}{\GG t}.
\end{eqnarray}
Or else, we may treat $x$ and $t$ in the same footing, that is 
set $\GG t \rightarrow P(x, t)\GG t$ and $\GG x \rightarrow P(x, t)
\GG x$ instead of only $\GG x \rightarrow P(x, t) \GG x$
and we get,
\begin{eqnarray}
\frac{\p P(x, t)}{\p t} = D P(x, t) \frac{\p^2 P(x, t)}{\p x^2},
\end{eqnarray}
both equations (23) and (25) being nonlinear equations. Henceforth 
we set $D = 1$. 
   
It should be noted that this does not mean any nonlinear equation 
can be obtained this way. The condition that for $\GG t = 0$ , 
$\GG x = 0$ both left and right side of equation (1) must match is 
quite a restriction. However, by making use of equation (7) one may
get many more equations by setting the source term as function of $x$,
$P$ and its derivatives or their combinations. This would be analogous
to adding terms to the Lagrangian arbitrarily in the conventional 
method of getting equations of motion. 

It is an interesting fact that if the master equation (1) is modified
in the first term of right hand side in the following way
\begin{eqnarray}
P_{\pm}(x, t + \GG t) = P_{\pm} (x \mp \GG x, t) (1 - P_+ \GG t) +
P_{\mp}(x \pm \GG x, t) a \GG t
\end{eqnarray}
we get a nonlinear form of the Dirac equation in one space dimension,
\begin{eqnarray}
\frac{\p P_+}{\p t} &=& - P_+^2 - v \frac{\p P_+}{\p x} + a P_- \\ 
\frac{\p P_-}{\p t} &=& - P_+ P_- + a P_+ + v \frac{\p P_-}{\p x}
\end{eqnarray}
and by iteration a nonlinear analogue of telegraphers' equation
results,
\begin{eqnarray}
\frac{\p^2 P_+}{\p t^2} - v^2 \frac{\p^2 P_+}{\p x^2} = 
- P_+^3 + P_+\frac{\p P_+}{\p t} + v P_+\frac{\p P_+}{\p x} + a^2 P_+ 
\end{eqnarray}

Further generalizations would be to consider $P$ as a complex multi-
component object and readers may amuse themselves by putting objects 
such as supersymmetric variables, Pauli and other matrices etc. as 
coefficients of $\GG x$ in equation (1).

\section{Group analysis}

Before we analyze the nonlinear equation (23) or (25) we would 
like to emphasize the important developments regarding the linear
diffusion equation (7). It is well known that this equation goes 
over to Burgers equation under the Cole-Hopf transformation. 
Burgers equation and similar integrable ones have been much studied
in recent decades for their importance in physical sciences for 
the existence of soliton like behaviour, infinite number of 
conservation laws as well as for their deep mathematical connections
to B\"acklund transformations, conformal invariance and so on. 

Equations of the form (23) have been analysed by Munier {\it etal}
\cite{Munier} and by Hill \cite{Hill} in detail. It is found that 
the nonlinear 
diffusion equation of the form
\begin{eqnarray}
\frac{\p\psi}{\p t} = \psi^2 \frac{\p^2 \psi}{\p P^2}
\end{eqnarray}
are equivalent to the classical diffusion equation for $P$,
\begin{eqnarray}
\frac{\p P}{\p t} = \frac{\p^2 P}{\p x^2}
\end{eqnarray}
if we introduce $x$ such that
\begin{eqnarray}
\psi (P, t) \equiv \frac{\p P}{\p x}
\end{eqnarray}
and every nonlinear diffusion equation of the form 
\begin{eqnarray}
\frac{\p P}{\p t} = \frac{\p}{\p x} \left [f(P) \frac{\p P}{\p x}
\right ]
\end{eqnarray}
can be transformed to the following equation with a simpler 
nonlinearity 
\begin{eqnarray}
f(P) \frac{\p \psi}{\p t} = \psi^2 \frac{\p^2 \psi}{\p P^2},
\end{eqnarray}
where $\psi(P, t)$ is the flux associated with equation (33). 
Hence, for this special case the analysis would be similar to that
of the linear diffusion equation. 

However, in general the simplest nonlinear equation we would get 
from the master equation by replacing $\GG x \rightarrow f(P)\GG x$
in equation (5) would be,
\begin{eqnarray}
\frac{\p P}{\p t} = f^2(P) \frac{\p^2 P}{\p x^2} 
\end{eqnarray}
as in equation (23) or
\begin{eqnarray}
\frac{\p P}{\p t} = f(P) \frac{\p^2 P}{\p x^2}
\end{eqnarray}
as in equation (25).

Now we proceed to analyse the properties of the solutions of 
(35) by means of one parameter groups as in 
\cite{Ovsiannikov,Bluman,Munier,Hill,Olver}. For 
the single dependent variable $P$ and for the two independent 
variables $x$ and $t$ we have one parameter groups of the form,
\begin{eqnarray}
x_1 &=& f(x, t, P, \epsilon) = x + \epsilon \xi (x, t, P) + 
O ({\epsilon}^2),
\nonumber\\
t_1 &=& g(x, t, P, \epsilon) = t + \epsilon \eta(x, t, P) + 
O ({\epsilon}^2), 
\nonumber\\
P_1 &=& h(x, t, P, \epsilon) = P + \epsilon \zeta(x, t, P)
+ O ({\epsilon}^2).
\end{eqnarray}
We follow the standard procedure 
\cite{Hill,Olver} to obtain the similarity 
variable and functional form of the solution by solving the first 
order partial differential equation
\begin{eqnarray}
\xi (x, t, P) \frac{\p P}{\p x} + \eta (x, t, P) \frac{\p P}{\p t}
= \zeta (x, t, P)
\end{eqnarray}
for known functions $\xi(x, t, P)$, $\eta(x, t, P)$ and 
$\zeta(x, t, P)$. Let
\begin{eqnarray}
{\bf v} = \xi (x, t, P) \frac{\p}{\p x} + \eta(x, t, P)\frac{\p}
{\p t} + \zeta (x, t, P) \frac{\p}{\p P}
\end{eqnarray}
be a vector field on the space $X\times U^{(2)}$, where coordinates
represent the independent variables, the dependent variables 
and the derivatives of the dependent variables up to order 2. All 
possible coefficient functions $\xi$, $\eta$, $\zeta$ are to be 
determined so that the one parameter group $\exp(\epsilon {\bf v})$,
thus obtained would be the symmetry group of the nonlinear equations 
(35) for the diffusion case and (29) for telegrapher's case. 

The determining equation for the symmetry group for the 
diffusion with non linearity, equation (35) is

\begin{tabular}{lllr}
\hline
\multicolumn{1}{c}{monomial} \qquad \qquad \qquad \qquad \qquad & 
\multicolumn{1}{c}{coefficients}
\qquad \qquad \qquad \qquad \qquad \qquad \qquad & 
\multicolumn{1}{c}{} \\ 
\hline

$\frac{\p^2 P}{\p x \p t} \frac{\p P}{\p x}$ &    $\eta_P = 0$ 
&   (A)\\
$\frac{\p^2 P}{\p x \p t}$ &  $\eta_x = 0$ &   (B)\\ 
${(\frac{\p P}{\p x})}^3$ &   $\xi_{P P} = 0$ &  (C)\\
${(\frac{\p P}{\p x})}^2 \frac{\p P}{\p t}$ &  $\eta_{P P}
= 0$ &  (D)\\
${(\frac{\p P}{\p x})}^2$ &  $(\zeta_P - 2 \xi_x)_P = 0$ &
(E)\\
${(\frac{\p P}{\p t})}^2$ &  $- \eta_P + \eta_P = 0$ & (F)\\
$(\frac{\p P}{\p x})(\frac{\p P}{\p t})$ &  $- \xi_P = 
- 2 \eta_{x P} f(P) - 3\xi_P$ &  (G)\\
$\frac{\p P}{\p x}$ &  $- \xi_t = f(P) (2 \zeta_{x P} - 
\xi_{x x})$ &  (H)\\
$\frac{\p P}{\p t}$ &   $\eta_t = f(P) \eta_{x x} + 2 \xi_{x} +
\frac{f'(P)\xi}{f(P)}$  &  (I)\\
$P^0$ &   $\zeta_t - f(P) \zeta_{x x} = 0$ &   (J)\\
\hline
\end{tabular}

\vspace{1cm}

where prime denotes differentiation with respect to the 
argument and subscript denote differentiation with respect to
the indicated variable. These equations turn out to be the 
same as those of the nonlinear diffusion equation of the 
form,
\begin{eqnarray}
\frac{\p P}{\p t} = \frac{\p}{\p x} \left [f(P) \frac{\p P}
{\p x}\right ]
\end{eqnarray}
considered in Hill \cite{Hill}. 

{}From $(A)$, $(B)$, and $(G)$ it is easily seen that 
\begin{eqnarray}
\xi = \xi(x, t) , \qquad \qquad \qquad \eta = \eta(t)
\end{eqnarray}
\begin{eqnarray}
\zeta_P = 2 \xi_x + r ,
\end{eqnarray}
where $r$ is a constant. So 
\begin{eqnarray}
\zeta_{P P} = 0.
\end{eqnarray}
{}From $(I)$ we get,
\begin{eqnarray}
\zeta = \frac{f(P)}{f'(P)} \left [2\xi_x - \eta_t \right ]
\end{eqnarray}
so that either
\begin{eqnarray}
2 \frac{\p \xi}{\p x} = \frac{\p \eta}{\p t} 
\end{eqnarray}
or
\begin{eqnarray}
{\left [\frac{f(P)}{f'(P)}\right ]}_{P P} = 0
\end{eqnarray}
that is,
\begin{eqnarray}
f(P) = a (P + b)^m
\end{eqnarray}
where $a$, $b$ and $m$ denote arbitrary constants. 
If equation (45) holds, then from $(h)$ and equation (44), we 
obtain, 
\begin{eqnarray}
\xi(x, t, P) &=& \beta + \gamma x 
\nonumber \\
\eta (x, t, P) &=& 2\theta + 2\gamma t 
\nonumber \\
\zeta(x, t, P) &=& 0
\end{eqnarray}
where, $\beta$, $\theta$ and $\gamma$ are arbitrary constants. 

Hence, the Lie algebra of infinitesimal symmetries of the equation
is spanned by the three vector fields,
\begin{eqnarray}
{\bf v_1} &=& \frac{\p}{\p x},  
\nonumber \\
{\bf v_2} &=& \frac{\p}{\p t}, 
\nonumber \\
{\bf v_3} &=& x \frac{\p}{\p x} + 2 t \frac{\p}{\p t},
\end{eqnarray}
and the commutation relations are given by,
\begin{eqnarray}
[{\bf v}_1, {\bf v}_2] = 0 , \qquad \qquad [{\bf v}_1, {\bf v}_3]
= {\bf v}_1 , \qquad \qquad [{\bf v}_2, {\bf v}_3] = 2 {\bf v}_2.
\end{eqnarray}

The one parameter groups $G_i$ generated by the ${\bf v}_i$ are 
given below. The entries give the transformed points $\exp(
\epsilon {\bf v}_i) (x, t, P) = (x_1, t_1, P_1)$. 
\begin{eqnarray}
G_1 : (x + \epsilon, t, P),
\nonumber \\
G_2 : (x, t + \epsilon, P),
\nonumber \\
G_3 : (e^{\epsilon} x, e^{2\epsilon} t, P).
\end{eqnarray}

Each group $G_i$ is a symmetry group and if $P = q (x, t)$ is a 
solution of our nonlinear diffusion equation, so are the  
functions,
\begin{eqnarray}
P^{(1)} &=& q (x - \epsilon, t),
\nonumber \\
P^{(2)} &=& q (x, t - \epsilon),
\nonumber \\
P^{(3)} &=& q (e^{-\epsilon} x, e^{-2\epsilon} t).
\end{eqnarray}

The groups we obtain are the same as those for equations (40) and
so is the similarity variable \cite{Hill} 
\begin{eqnarray}
\omega = \frac{x + \A}{(t + \beta)^{1/2}}.
\end{eqnarray}
However, the functional form
\begin{eqnarray}
P = s (\omega)
\end{eqnarray}
of the solution satisfies the ordinary differential equation
\begin{eqnarray}
2 f(s) \frac{d^2 s}{d \omega^2} + \omega \frac{d s}{d \omega}
= 0, 
\end{eqnarray}
whereas that corresponding to equation (40) is given by, 
\begin{eqnarray}
2 f(s) \frac{d^2 s}{d \omega^2} + 2 \frac{d f(s)}{d s} 
\left (\frac{d s}{d \omega}\right )^2 + 
\omega \frac{d s}{d \omega} = 0.
\end{eqnarray}
In the case $f(P)$ is given by equation (46),
\begin{eqnarray}
\zeta = (\frac{P + b}{m}) \left [2 \frac{\p \xi}{\p x} - 
\frac{\p \eta} {\p t}\right ] 
\end{eqnarray}
and for the time derivative of $\xi$ we get,
\begin{eqnarray}
\frac{\p \xi}{\p t} = f(P) \left [1 - \frac{4}{m}\right ] 
\frac{\p^2 \xi} {\p x^2}. 
\end{eqnarray}
while substituting (57) into $(J)$ and using (58) gives,
\begin{eqnarray}
\eta_{t t} = - \frac{8}{m} \xi_{x x x}.
\end{eqnarray}
So there are two possibilities arising out of equation (58), 
either for all constants $m$ ,
\begin{eqnarray}
\frac{\p \xi}{\p t} = \frac{\p^2 \xi}{\p x^2} = \frac{
\p^2 \eta}{\p t^2} = 0,
\end{eqnarray}
or for $m = 4$ ,
\begin{eqnarray}
\frac{\p \xi}{\p t} = \frac{\p^3 \xi}{\p x^3} = \frac{\p^2 \eta}
{\p t^2} = 0.
\end{eqnarray}
Thus for all $m$ we have,
\begin{eqnarray}
\xi(x, t, P) &=& \mu + \sigma x, 
\nonumber \\
\eta (x, t, P) &=& \nu + \rho t, 
\nonumber \\
\zeta (x, t, P) &=& \left (\frac{P + b}{m}\right ) (2 \sigma - \rho), 
\end{eqnarray}
where $\mu$, $\nu$, $\sigma$ and $\rho$ are arbitrary constants 
and the infinitesimal symmetries are spanned by four vector fields
\begin{eqnarray}
{\bf v}_1 &=& \frac{\p}{\p x},
\nonumber \\
{\bf v}_2 &=& \frac{\p}{\p t},
\nonumber \\
{\bf v}_3 &=& x \frac{\p}{\p x} + \frac{2}{m} (P + b) 
\frac{\p}{\p P},
\nonumber \\
{\bf v}_4 &=& t \frac{\p}{\p t} - \frac{(P + b)}{m} \frac{\p}
{\p P},
\end{eqnarray}
and the commutation relations are given by,

\begin{eqnarray}
[{\bf v}_1, {\bf v}_2] = [{\bf v}_1, {\bf v}_4] = [{\bf v}_2, 
{\bf v}_3] = [{\bf v}_3, {\bf v}_4] = 0, \qquad [{\bf v}_1, 
{\bf v}_3] = {\bf v}_1, \qquad [{\bf v}_2, {\bf v}_4] = {\bf v}_2.
\end{eqnarray}

The one parameter groups $G_i$ generated by the ${\bf v}_i$ 
are,
\begin{eqnarray}
G_1 &:& (x + \epsilon, t, P), 
\nonumber \\
G_2 &:& (x, t + \epsilon, P),
\nonumber \\
G_3 &:& (e^{\epsilon} x, t, (P + b) e^{\frac{2\epsilon}{m}}),
\nonumber \\
G_4 &:& (x, e^{\epsilon t}, (P + b) e^{- \frac{\epsilon}
{m}}),
\end{eqnarray}
and if $P = y(x, t)$ is a solution to our non linear diffusion 
equation, so are the functions
\begin{eqnarray}
P^{(1)} &=& y (x - \epsilon, t, P),
\nonumber \\
P^{(2)} &=& y (x, t - \epsilon, P),
\nonumber \\
P^{(3)} &=& y (e^{-\epsilon} x, t, (P - b) e^{-\frac{2\epsilon}
{m}}),
\nonumber \\
P^{(4)} &=& y ( x, e^{-\epsilon} t, (P - b) e^{\frac{\epsilon}
{m}}).
\end{eqnarray}
The similarity variable in this case is given by,
\begin{eqnarray}
\omega = \frac{x + \frac{\mu}{\sigma}}{{{(t + \frac{\nu}
{\rho})}}^{\frac{\sigma}{\rho}}} 
\end{eqnarray}
and the functional form of the solution is,
\begin{eqnarray}
P = {\left (t + \frac{\nu}{\rho}\right )}^{(\frac{2 \sigma}{m \rho} 
- 1)}
s (\omega) - b.
\end{eqnarray}

Now for the nonlinear form of the telegrapher's equation
(29), arising out of the nonlinear Dirac equation (27), the
independent determining equations of the symmetry group are
given below.
\vspace{2cm}

\begin{tabular}{lllr}
\hline
\multicolumn{1}{c}{monomial} 
\qquad \qquad \qquad \qquad &
\multicolumn{1}{c}{coefficient} 
\qquad \qquad \qquad \qquad &
\multicolumn{1}{c}{} \\
\hline
$\frac{\p^2 P_+}{\p t^2}  \frac{\p P_+}{\p t}$ &  $\eta_{P{_+}} 
= 0$ &   (a) \\
$\frac{\p^2 P_+}{\p x \p t}  \frac{\p P_+}{\p t}$ & 
$\xi_{P_{+}} = 0$ &   (b) \\
$\frac{\p^2 P_+}{\p x \p t}$ &  $\xi_t = v^2 \eta_x$ & 
(c) \\
$\frac{\p^2 P_+}{\p t^2}$ &  $\eta_t = \xi_x$ &  (d) \\
$\frac{\p P_+}{\p t})$ &  $2\zeta_{t P_{+}} - \eta_{t t} + 
v^2 \eta_{x x} + P_+(\zeta_{P_{+}} - 2\xi_x) - $\\
& $ P_+v\eta_x - P_+
\eta_x = 0$ &  (e) \\     
$\frac{\p P_+}{\p x}$ &  $\xi_{t t} + v^2(2\zeta_{x P_{+}} - \xi_{x x}) +
 \zeta + $\\ 
& $P_+v(\zeta_{P_{+}} - 2\xi_x) + P_+(\zeta_{P_{+}} - \xi_x) + $\\ 
& $ vP_+(\zeta_{P_{+}} - \xi_x)  
= 0$ &  (f)  \\ 
$(P_+)^0$ &  $\zeta_{t t} - v^2 \zeta_{x x} - P_+^3(\zeta_{P_{+}} - 2
\xi_x) + $\\
& $ P_+a^2(\zeta_{P_{+}} - 2\xi_x) + 3{P_+}^2\zeta + P_+\zeta_x + $\\ 
&  $v\zeta_{xP_{+}} + a^2\zeta = 0$ &  (g) \\ 
\hline
\end{tabular}
\vspace{2cm}

The solutions are given by,
\begin{eqnarray}
\xi (x, t, P_+) &=& Av^2 t + B , 
\nonumber \\
\eta (x, t, P_+) &=& Ax + E, 
\nonumber \\
\zeta(x, t, P_+) &=& 0,
\end{eqnarray}
where, $A$, $B$, and $E$ are arbitrary constants and 
the infinitesimal 
symmetries are spanned by the three vector fields 
\begin{eqnarray}
{\bf v}_1 &=& \frac{\p}{\p x} , \qquad space \  translation,
\nonumber \\
{\bf v}_2 &=& \frac{\p}{\p t} , \qquad time \ translation, 
\nonumber \\
{\bf v}_3 &=& v^2 t \frac{\p}{\p x} + x\frac{\p}{\p t} ,
\qquad hyperbolic \ "rotation"\ in \ x, \ t \ space, 
\end{eqnarray}
with the commutation relations,
\begin{eqnarray}
[{\bf v}_1, {\bf v}_2] =0,   \quad
[{\bf v}_1, {\bf v}_3] = {\bf v}_2, \quad
[{\bf v}_2, {\bf v}_3] = v^2 {\bf v}_1 
\end{eqnarray}

The one parameter groups $G_i$ generated by the ${\bf v}_i$ are,
\begin{eqnarray}
G_1 &:& (x + \epsilon, t, P_+), 
\nonumber \\
G_2 &:& (x, t + \epsilon, P_+), 
\nonumber \\
G_3 &:& (x + v^2 \epsilon t, t + \epsilon x, P_+).
\end{eqnarray}

This implies that if $P_+ = z (x, t)$ is a solution to equation
(29), so are the functions

\begin{eqnarray}
P_{+ 1} &=& z(x - \epsilon, t),
\nonumber \\
P_{+ 2} &=& z(x, t - \epsilon),
\nonumber \\
P_{+ 3} &=& z(x - v^2 \epsilon t, t - \epsilon x),
\end{eqnarray}
where $\epsilon$ is any real number.

In order to compare the above vector fields of equation (70)
with those of the linear second order form of the telegrapher's
equation (3), we have the corresponging independent determining
equations of the symmetry group:
\vspace{2cm}

\begin{tabular}{lllr}
\hline
\multicolumn{1}{c}{monomial}
\qquad \qquad \qquad \qquad  &
\multicolumn{1}{c}{coefficient}
\qquad \qquad \qquad \qquad &
\multicolumn{1}{c}{} \\
\hline
$\frac{\p^2 P_+}{\p x^2}  \frac{\p P_+}{\p x}$ &  $\xi_{P_+}
= 0$ &  (P) \\
$\frac{\p^2 P_+}{\p x^2} \frac{\p P_+}{\p t}$ &  $\eta_{P_+}
= 0$ &  (Q) \\
$\frac{\p^2 P_+}{\p x^2}$ & $\xi_x = \eta_t$ &  (R)  \\
$\frac{\p^2 P_+}{\p x \p t}$ &  $\xi_t = v^2 \eta_x$ &  (S)
\\
$(\frac{\p P_+}{p t})^2$ &  $\zeta_{P_+ P_+} - 2\eta_{t P_+}
+ 4a\eta_{P_+} = 0$ &  (T) \\
$\frac{\p P_+}{\p x}$ & $\xi_{t t} - v^2\xi_{x x} + 
2v^2\zeta_{x P_+} + 2a\zeta_t = 0$ &  (U) \\
$\frac{\p P_+}{\p t}$ & $\eta_{t t} - v^2\eta_{x x} - 
2\zeta_{t P_+} - 2a\eta_t = 0$ &  (V)\\
$(P_+)^0$ &  $\zeta_{t t} - v^2\zeta_{x x} + 2a\zeta_t 
= 0$ &  (W) \\
\hline
\end{tabular}
\vspace{2cm}

The solutions are given by,
\begin{eqnarray}
\xi (x,t,P_+) &=& Kv^2 t + L,
\nonumber \\
\eta (x,t,P_+) &=& Kx + M,
\nonumber \\
\zeta (x,t,P_+) &=& -KaxP_+ + NP_+,  
\end{eqnarray}
where K,L,M, and N are arbitrary constants. The
infinitesimal symmetries are spanned by the four 
vector fields
\begin{eqnarray}
{\bf v}_1 &=& \frac{\p}{\p x} ,
\nonumber \\
{\bf v}_2 &=& \frac{\p}{\p t},
\nonumber \\
{\bf v}_3 &=& P_+ \frac{\p}{\p P_+},
\nonumber \\
{\bf v}_4 &=& v^2 t\frac{\p}{\p x} + x\frac{\p}{\p t} - 
axP_+ \frac{\p}{\p P_+},
\end{eqnarray}
with commutation relations
\begin{eqnarray}
{[{\bf v}_1, {\bf v}_2]  = [{\bf v}_1, {\bf v}_3] = 0,\quad
[{\bf v}_1, {\bf v}_4] = {\bf v}_2 - a{\bf v}_3,}
\nonumber \\
{[{\bf v}_2, {\bf v}_3] = 0}, \quad 
{[{\bf v}_2, {\bf v}_4] = v^2{\bf v}_1},\quad
{[{\bf v}_3, {\bf v}_4] = 0}.
\end{eqnarray}

We have ignored the obvious infinite dimensional
subalgebras in the above analysis.

\section{Conclusion}

The main objective of this paper was to extend the method of 
deducing some fundamental linear partial differential equations 
of physics from a consideration of stochastic arguments to the 
nonlinear case. We saw that this could be achieved in a very 
simple way by modifying the master equation to obtain "nonlinear 
diffusion" equation, a "nonlinear Dirac equation" in 1 + 1
 dimensions
and the corresponding "non linear telegrapher's equation".
 As a preliminary 
step towards the analysis of the properties of the solutions,
we have considered the group classification problem of the first 
and the last one by means of one parameter groups. The infinitesimal
symmetry group of the nonlinear telegrapher's equation 
is spanned by a vector field 
corresponding to a "hyperbolic rotation" of $x$ and $t$. For 
our type of diffusion equation, though the group structure 
is similar to that of standard nonlinear diffusion equation, 
the ordinary differential equations 
obtained is different and the results are similar when $m = 4$ in 
our case, but $m = - \frac{4}{3}$ in the standard case ($m$ 
being the highest power of the dependent variable in coefficient
to the $\frac{\p^2}{\p x^2}$ term in the nonlinear diffusion 
equation). The physical applications of this equation has been
widely studied in the context of gas dynamics and plasma physics etc.
We expect the other two equations to have 
similar important applications in physics with rich mathematical 
structure and we leave it for future study.However, as a comparison
of equations (70) and (75) shows  one does 
see explicitly which symmetries gets broken when the equation 
is modified.

\noindent
{\bf Acknowledgments}  \medskip \newline

The author is grateful to Prof. J. Ehlers for providing 
academic facilities at the Albert Einstein Institut, Potsdam,
Germany, where this work was carried out. The author also wishes
to thank S. Mahapatra for comments on the manuscript.    

%
%

\renewcommand{\arraystretch}{1}

\newcommand{\NP}[3]{{ Nucl. Phys.} {\bf #1} {(19#2)} {#3}}
\newcommand{\PL}[3]{{ Phys. Lett.} {\bf #1} {(19#2)} {#3}}
\newcommand{\PRL}[3]{{ Phys. Rev. Lett.} {\bf #1} {(19#2)} {#3}}
\newcommand{\PR}[3]{{ Phys. Rev.} {\bf #1} {(19#2)} {#3}}
\newcommand{\IJ}[3]{{ Int. Jour. Mod. Phys.} {\bf #1} {(19#2)}
  {#3}}
\newcommand{\CMP}[3]{{ Comm. Math. Phys.} {\bf #1} {(19#2)} {#3}}
\newcommand{\PRp} [3]{{ Phys. Rep.} {\bf #1} {(19#2)} {#3}}

\end{document}